\documentclass[prd,
preprint,nofootinbib%
 ,secnumarabic%
,amssymb, amsmath,mathrsfs,nobibnotes,aps,11pt]{revtex4}
\usepackage[utf8]{inputenc}
\usepackage{epsfig}
\usepackage{color}
\usepackage{ulem}
\usepackage{graphicx}
\usepackage{float}
\usepackage{latexsym}
\usepackage{amscd}
\usepackage{fancyhdr}
\usepackage{setspace}
\usepackage{subfigure}
\usepackage{braket}
\usepackage{slashed}

\usepackage{tabularx}
\usepackage[colorlinks=true,citecolor=blue,linkcolor=blue]{hyperref}%
\begin{document}

\title{Viscous dark matter and 21 cm cosmology}

\author{Jitesh R. Bhatt$^{1,}$\footnote{ jeet@prl.res.in}}
\author{Arvind Kumar Mishra$^{1,2}$\footnote{arvind@@prl.res.in}}
\author{Alekha C. Nayak$^{1,}$\footnote{ alekha@@prl.res.in}}
\affiliation{$^1$Theoretical Physics Division, Physical Research Laboratory,
	Navrangpura, Ahmedabad, 380009, India}
\affiliation{$^2$Indian Institute of Technology Gandhinagar, Palaj, Gandhinagar, 382424, India }		
\def\be{\begin{equation}}
\def\ee{\end{equation}}
\def\al{\alpha}
\def\bea{\begin{eqnarray}}
\def\eea{\end{eqnarray}}


\begin{abstract}
The EDGES experiment has detected the global absorption signal of 21 cm line at $z\sim17$ in cosmic dawn era and reported its amplitude larger than the standard cosmological prediction. One of the possible explanation requires that the baryons were much cooler than the standard scenario.
This requires an
interaction between the dark and baryonic sectors with some appropriate cross-section, $ \hat{\sigma} $ .
In this work, we examine the role that dissipative effects of cosmic fluid might play in influencing the 21 cm signal. We show that the presence of viscous dissipation of dark matter can significantly affect the energy transfer between the baryonic and dark matter fluids. It is demonstrated that the inclusion of the dissipative mechanism in the dark sector, strongly modify the earlier constraints on dark matter mass and $ \hat{\sigma} $ obtain from EDGES observation. Further, we argue that EDGES absorption signal can put an independent bound on dark matter viscosity which is many order of magnitude larger than the maximum viscosity allowed by the structure formation.  
\end{abstract}
\maketitle

\section{Introduction}
Recently EDGES has reported an absorption signal of 21 cm line with an amplitude of $0.5$ degree Kelvin at the redshift $z\sim17$ \cite{Bowman:2018yin}. This signal was reported in the cosmic dawn era $15 \leq  z \leq 20$, when the Universe was in its coolest phase and the star formation started for the first time in the cosmic history of the Universe. Later, as the star formation proceeds, the ultraviolet radiation emitted from the stars heat up the baryonic matter and the spin temperature becomes larger than the CMB temperature and hence 21 cm signal goes to zero below $z\sim 15$.
The amplitude of the reported absorption signal is approximately two times larger than the standard cosmological prediction. 
The absorption dip indicates that either the baryons were much cooler or the photons were much hotter than what one would predict from the standard cosmological scenario \cite{Barkana:2018lgd,Barkana:2018qrx}. 

One of the promising ways to reduce the baryon temperature could be an interaction between the baryons and the dark matter (DM)  \cite{Barkana:2018lgd,Barkana:2018qrx} because, at the cosmic dawn era, the DM was much colder than the baryons. For a sufficient cooling of baryons at cosmic dawn era, the preferable form of DM-baryon scattering is assumed as $\sigma=\hat \sigma v^{-4}_{\mathrm{rel}}$, where $v_{\mathrm{rel}}$ represents the relative velocity between DM and baryons \cite{Tashiro:2014tsa,Munoz:2015bca,Munoz:2018pzp}. It has been shown that in order to explain the EDGES observation due to DM-baryon interaction, the DM mass should be sub GeV scale,  $m_{\chi} \leq 1$ GeV \cite{Barkana:2018lgd,Barkana:2018qrx}.
In the standard lambda cold dark matter ($\Lambda $CDM) model, 
the widely favored  DM candidate, weakly interacting massive relic particles (WIMPs),  which fits with the Planck observational constraint has a mass range of few GeV to TeV. However, the non-observation of WIMP at direct detection experiments XENON100 \cite{xenon}, LUX \cite{daSilva:2017swg} and at LHC excluded large parameter space of DM mass.  The direct detect detection experiments are highly sensitive around DM masses 30 GeV and less sensitive below DM mass below 10 GeV due to small nuclear recoil energy.
Since direct detection experiments are not sensitive to such low mass of DM  particles (i.e. $m_{\chi} \leq 1$ GeV) that explain the EDGES observation, hence it evades the direct detection DM mass constraints.
The other possible alternative explanations of EDGES observation include the emission of 21-cm axion \cite{Lambiase:2018lhs,Houston:2018vrf,Auriol:2018ovo}, excess of early radio background  \cite{Yang:2018gjd} and early dark energy \cite{Hill:2018lfx} etc.

The reported EDGES signal in the cosmic dawn era provides the unique opportunity to understand the microscopic properties of the dark matter.
In Refs.\cite{Liu:2018uzy,DAmico:2018sxd}, it has been shown that dark matter annihilation, decay can cause energy injection in the cosmic medium and thus EDGES observation used to obtain constraints on dark matter annihilation and decay rates. Later, in Ref.\cite{Lopez-Honorez:2018ipk} it has been argued that the interaction between the DM and lighter degree of freedom, can delay the 21 cm absorption signal due to collisional damping and put a strong bound on
DM-interaction with the light Standard model particles. Further, in Ref.\cite{Kovetz:2018zes} EDGES observation has been used to constrain on kinematic mixing strength between the photon and hidden photon in ultra-light hidden photon DM model.

In the standard cosmology, dark matter is assumed to be ideal but if dark matter is viscous then it will change the standard cosmological evolution history. The viscosity of the DM is related to the microscopic properties of the DM particles.
In Ref.\cite{Atreya:2017pny}, two of us have argued that the Self Interacting Dark Matter (SIDM) that solves the small-scale issues of collisionless cold
 dark matter paradigm may produce the viscosity (both bulk and shear) and hence contribute to cosmic dissipation. The
 viscous effects of SIDM are sufficient enough to produce the late time accelerated expansion \cite{Atreya:2017pny}
 and also explain the late time cosmology without any need of separate dark energy component
 \cite{Atreya:2018iom}.
 Also in the Ref.\cite{Mathews:2008hk}, it has been shown that the decay of cold dark matter into relativistic particles can produce the bulk viscosity. The phenomenological implications of viscous dark matter are very rich and have been applied in the different aspects of the cosmology. In  the earlier works, it has been also argued that if dark matter has sufficient bulk viscosity then it can produce inflation like behaviour \cite{Padmanabhan:1987dg, Gron:1990ew, Cheng:1991uu, Zimdahl:1996ka} and  explain the present observed cosmic acceleration \cite{Fabris:2005ts,Avelino:2008ph,Das:2008mj,  Piattella:2011bs, Velten:2011bg, Gagnon:2011id, Mohan:2017poq, Cruz:2018yrr,Li:2009mf,Barbosa:2015ndx,Floerchinger:2014jsa}. Some recent discussions on viscous cosmology can be found in Refs.\cite{Brevik:2017msy, Anand:2017wsj, Cai:2017buj,Anand:2017ktp,Lu:2018smr,Brevik:2019yma}.

In this work, we study the dissipative effect of DM along with the DM-gas interaction in the light of reported absorption signal. Since the shear viscosity is
severely constrained by the homogeneity and isotropy of the Universe \cite{Velten:2013pra}, hence we neglect the shear viscosity and assume only the bulk viscosity in the cosmic fluid.
Here we focus on the dissipative effect of the dark sector from DM viscosity that may arise from the DM self-interaction, dark matter decay etc. Further,
the presence of the dissipative effect of a DM fluid produces the entropy \cite{Weinberg:1972kfs} throughout the cosmic evolution history and heats up the dark matter fluid. We consider the phenomenological form of the dark matter bulk viscosity as $\zeta_{\chi} = \zeta_{0}\left( \frac{\rho_{\chi}(z)}{\rho_{\chi}(0)}\right)^\gamma$ and calculate the entropy production using the FLRW metric in the expanding universe. Then, we set up the temperature evolution equations for the viscous dark matter and gas in the presence of DM-gas interaction. Later, we check the dependency of DM dissipation, DM mass and DM-gas interaction on the DM and gas temperature throughout the cosmic evolution. 

We find that the low DM viscosity does not generate sufficient entropy to change the DM temperature and the DM fluid behaves like an ideal fluid.  But, if the viscosity of the DM is sufficiently large then it increases DM temperature and also the gas temperature through the DM-baryon scattering at the low redshift. The cooling of the gas can be done by either increasing the DM-gas interaction or decreasing the DM mass. Hence, in
order to explain the EDGES absorption signal in presence of viscous DM scenario, we need larger  DM-gas cross-section and smaller DM mass in comparison with the ideal DM fluid. This allows us to 
put the new limits on the DM mass and the DM-baryon interaction cross-section. Further, using the limits on the $\hat{\sigma}$ from the CMB, we constrain the DM viscosity. We find that the EDGES observation can allow a large viscosity in the dark sector, which is approximately $10^{4}$ to $10^{6}$ orders  larger in comparison with the limit obtained from the structure formation.

The organization of our paper is as follows: In Section \ref{sec:21cmsignal}, we will discuss the basic of 21 cm signal and also observed EDGES measurements. In Section \ref{sec:viscdm}, we will derive the evolution equation of dark matter temperature in presence of the dark matter viscosity.  Assuming the DM-baryon interaction and DM viscosity, we will set up the temperature evolution equations of dark matter and baryons in Section \ref{sec:darkheating}. Later in Section \ref{sec:results}, we show our results and put constraints on the DM viscosity, DM mass and DM-baryon scattering using the EDGES observational signal. Finally,  we conclude our work in Section \ref{sec4}. 
\section{Standard model of 21 signal and EDGES observation}
\label{sec:21cmsignal}
In this Section, we will discuss the basics of the 21 cm signal using the standard cosmological model. Later, we also discuss the recently observed 21 cm global absorption signal by the EDGES experiment.

The 21 cm absorption/emission line emits from the spin slip transition between two hyperfine states, singlet ($F=0$) and triplet ($F=1$) of the ground level of the hydrogen atom.  The relative population of triplet ($n_1$) and singlet  ($n_0$) of hyperfine splitting is characterized by the spin temperature $T_{S}$. The spin temperature is given by
\begin{equation}
	\frac{n_{1}}{n_{0}} = \frac{g_{1}}{g_{0}}~e^{-\frac{ E_{21}}{T_{S}}} \simeq 3\left( 1 - \frac{ E_{21}}{T_{S}}\right)
\end{equation}
where $ E_{21} $ is the energy of 21 cm line.
Throughout the cosmic history of the Universe, the $ T_{S} $ is determined by
three competing mechanisms: CMB radiation, collisions and late time Lyman-$\alpha$ radiation.  In the equilibrium, the spin temperature is given as  
\begin{equation}
	T_{S} = \frac{T_{C} + y_{c}T_{G} + y_{w}T_{Ly\alpha}}{1 + y_{c} + y_{w} }
	\label{eq:spintem}
\end{equation}
where $ T_{C} $ and $ T_{G} $ represents the CMB  and kinetic temperature of gas. The temperature, $ T_{Ly\alpha} $ is defined through the detailed balance equation. The other quantities  $ y_{c} $ and $  y_{w} $ is defined as
\begin{equation}
	y_{c} = \frac{T_{S}}{T_{G}}\frac{P^{c}_{01} }{A_{10} }, \ \  y_{w} = \frac{T_{S}}{T_{Ly\alpha}}\frac{P^{w}_{01} }{A_{10} }.
\end{equation}
Here $ P^{c}_{01} $ and $ P^{w}_{01} $ are the probabilities of the ground and exited state through the collisions and Lyman-alpha radiation. The $A_{10} $ is the rate of the spontaneous decay, which value is $ \sim 2.9\times 10^{-15} $ sec$^{-1}$\cite{Barkana:2018lgd}.

In the standard cosmology, after recombination, the gas kinetically decouples from cosmic microwave background at $z\approx 1100$.  The gas remains thermally couple with CMB vis Compton scattering off the residual electrons and $T_{S} =T_{G}$ until $z\sim 150$. Below redshift $z<150$, due to decreasing electron fraction, the Thomson scattering between the CMB and gas becomes unimportant and gas get thermally decoupled from the CMB radiation and cools adiabatically for the rest of cosmic evolution. At the smaller redshift $z\sim 100$, the collision between the
$H-H, H-e$ in the gas becomes prominent and hence $T_{S}$  couples gas temperature,
and $T_{S}$ follow the gas temperature, i.e. $T_{S} =T_{G}$. But at a later time due to decreasing the number density of the gas, the collisions between the gas become less effective and  $T_{S}$ no longer hold the gas temperature $T_{G}$. At this time gas absorbed/emitted the CMB radiation efficiently and spin temperature becomes CMB temperature. Finally at  $z<20$, stellar Lyman $\alpha$ couples spin temperature with gas via Wouthuysen-Field effect such that $T_S$ equals to $T_G$. In this paper, for simplicity, we will use words gas and baryon interchangeably unless otherwise specified.

The intensity of the observed signal is quantified in terms of brightness temperature \cite{Barkana:2018lgd,Barkana:2018qrx}
\begin{equation}
	T_{21}= \frac{1}{1+z}(T_{S}-T_{C})(1-\exp^{-\tau})
	\label{eq:brightness}
\end{equation}
Here  $\tau$ is the optical depth, which is given by
\begin{equation}
	\tau \approx \frac{3\lambda^{2}_{21}A_{10}n_{H}}{16T_{S}H(z)}~,
	\label{eq:optdepth}
\end{equation}
where $ \lambda_{21} $ is the wavelength of $21$cm at rest and $ n_{H} $ is hydrogen number density. Assuming $T_S=T_G$, the standard astrophysics prediction for brightness temperature at  $z \simeq 17 $  is 
\bea
T_{21}\geq -220 ~ mK
\eea
and corresponding gas temperature $6.8$ K \cite{Barkana:2018qrx}.  Recently, EDGES \cite{Bowman:2018yin} measured global 21 cm absorption signal from the cosmic dawn era and absorption centered around frequency $\nu \approx 78$ MHz (redshift $z\approx 17$). EDGES brightness measurement at $z \simeq 17 $  with $99\%$ confidence interval is given by 
\bea
T_{21}^{EDGES}\approx -500^{+200}_{-300}~ mK
\eea
In the optimal condition, this corresponds to gas temperature: $3.26^{+1.94}_{-1.58}~ K$ \cite{Barkana:2018qrx}. It is evident from the Eq.(\ref{eq:brightness}) that any mechanism that reduces the brightness temperature can explain the EDGES signal. It have been argued that the excess absorption deep of the EDGES observation can be explained by either decreasing $T_S$ via elastic scattering of dark matter with gas \cite{Barkana:2018lgd,Barkana:2018qrx} or increasing the $T_C$ via increasing the photon density, i.e. conversion of axion to photon \cite{Moroi:2018vci} during the cosmic dawn. 

Furthermore, the presence of the dissipative mechanism may increase the brightness temperature (i.e. decrease the strength of the absorption signal) by heating the gas.
For example, the dark matter annihilation/decay can heat up the gas temperature $T_G$ by injecting energy into the intergalactic medium which could possibly erase the absorption signal \cite{Liu:2018uzy,DAmico:2018sxd}.
Here, we consider the heating effect coming due to the viscosity of dark matter and put the limit on DM viscosity, dark matter mass and DM-baryon cross-section using the EDGES absorption signal in next upcoming Sections.   
\section{viscous dark matter and Temperature evolution }
\label{sec:viscdm}
In this Section, we derive the temperature evolution of dark matter particles in the presence of dark matter viscosity.
Using the first and second law of thermodynamics, we get the temperature evolution of viscous dark matter particle as 
\begin{equation}
\frac{dT_{\chi}}{dz}=2 \frac{T_{\chi}}{1+z} - \frac{2}{3(1+z)H}\frac{m_{\chi}}{\rho_{\chi}}\frac{dQ_{\mathrm{v}}}{dVdt}~.
\label{eq:dmtemp}
\end{equation}
Where $T_{\chi}$, $m_{\chi}$, $\rho_{\chi}$ represents the temperature, mass, energy density of the dark matter and $H$ is the Hubble expansion rate. In the above equation, the first term corresponds to the Hubble dilution and second term of account for the heating effect which is coming from the DM dissipation. 
The presence of viscosity in the dark matter leads to entropy generation which heats the dark matter.
The entropy production by the imperfect fluid in expanding Universe is calculated in Ref.\cite{Weinberg:1972kfs}. We apply this formalism for dark matter and find the entropy production per unit volume due to bulk viscous dark matter using the FLRW metric as
	\begin{equation}
\nabla_{\mu}S^{\mu} = \frac{\zeta_{\chi}}{T_{\chi}}\bigg( \nabla_{\mu}u^{\mu}\bigg)^{2},
\end{equation}
where $S^{\mu} $ is entropy four vector, given by
\begin{equation}
S^{\mu} = n_{\chi} s_{\chi}u^{\mu}~.
	\end{equation}
Where $s_{\chi} $,  $n_{\chi}$ and $u^{\mu}$ represents the entropy per unit particle, number density and four-velocity of the dark matter respectively. Due to dark matter viscosity, the heat energy per unit time per unit volume generated by viscous dark matter fluid is given by the second law of thermodynamics
	\begin{equation}
\frac{dQ_{\mathrm{v}}}{dVdt} =   T_{\chi}\nabla_{\mu}S^{\mu}= \zeta_{\chi}\bigg( \nabla_{\mu}u^{\mu}\bigg)^{2}
\label{new_vis}
	\end{equation}
In the comoving frame, the above Eq.\eqref{new_vis} is  rewritten as 
	\begin{equation}
\frac{dQ_{\mathrm{v}}}{dVdt} =   \zeta_{\chi}\left(3H \right)^2~. 
	\label{eq:heating} 
	\end{equation}
Hence in presence of DM viscosity, the dark matter temperature evolution is obtained by applying Eq.(\ref{eq:heating}) into the Eq.\eqref{eq:dmtemp} as
	\begin{equation}
\frac{dT_{\chi}}{dz}=2 \frac{T_{\chi}}{1+z} - \frac{6}{(1+z)H}\left[ \frac{m_{\chi}\zeta_{\chi}H^{2}}{\rho_{\chi}}\right] 
	\label{eq:viscous}
	\end{equation}
The above equation shows that the presence of DM viscosity modifies the DM temperature evolution and because of $\zeta_{\chi}>0$ the dark matter temperature will always increase throughout the cosmic evolution.
 
To study the dissipative effect of dark matter, we consider a phenomenological choice of bulk viscosity as \cite{Velten:2013pra}
	\begin{equation}
\zeta_{\chi} = \zeta_{0}\left( \frac{\rho_{\chi}(z)}{\rho_{\chi_{0}}}\right)^{\gamma}~~,
	\label{eq:viscdef} 
	\end{equation}
where $ \rho_{\chi}(z)$  and $\rho_{{\chi}{0}}$ represents the energy density of dark matter at redshift $z$ and presents. Here $ \gamma $ and $ \zeta_{0} $ represent the free parameters of dark matter viscosity. In this paper, we will not discuss the specific mechanism for bulk viscosity production but for sake of understanding, one can assume that the source of viscosity may be DM self-interaction, DM decay.   

In our viscous DM model, we are considering that Universe  consists of radiation ($R$), baryon ($B$), viscous cold dark matter ($\chi$) and cosmological constant ($\Lambda$).
The energy density of the viscous dark matter, $ \rho_{\chi}(z) $ can be calculated by applying the continuity equation for viscous DM component, in term of redshift \cite{Velten:2013pra}
	\begin{equation}
\frac{d\Omega_{\chi}(z)}{dz} - \frac{3}{1+z}\Omega_{\chi}(z) + \frac{\bar{\zeta}}{1+z}\left( \frac{\Omega_{\chi}(z)}{\Omega_{\chi}(0)}\right)^{\gamma}\left[ \Omega_{R0}(1+z)^4 + \Omega_{B0}(1+z)^3 + \Omega_{\mathrm{\chi}}(z) + \Omega_{\Lambda}\right]^{1/2} = 0.
	\label{eq:viscevol}
	\end{equation}
Here $ \Omega_{\mathrm{i}}(z) = \frac{4\pi G \rho_{\mathrm{i}}(z)}{3H^{2}_{0}} $, where $ i=R, B, \chi,\Lambda $ and  $ \Omega_{\chi}(0)=\Omega_{\chi_{0}}$. 
In the Eq.(\ref{eq:viscevol}), the subscript, $0$ represents the present values of the respective quantities and their values are taken from the Ref. \cite{Ade:2015xua}. The other quantity $\bar{\zeta} $ represents the dimensionless viscosity parameter which is related with the viscosity coefficient  $\zeta_{0} $ by $ \bar{\zeta} = \frac{24\pi G\zeta_{0}}{H_{0}}$.
Here the initial condition of $\Omega_{\chi}$ is given by its present value i.e. $  \Omega_{\chi}(0) = \Omega_{\chi_{0}}$ 
Thus in the viscous dark matter model, the Hubble expansion rate is given by
	\begin{equation}
H = H_{0}\bigg[ \Omega_{R0}(1+z)^4 + \Omega_{B0}(1+z)^3 + \Omega_{\chi}(z) + \Omega_{\Lambda} \bigg]^{1/2} 
	\label{eq:hubble}
	\end{equation}
Here we see that the viscosity does not  affect only the DM temperature but for sufficiently large viscosity, it may also affect the background expansion of the Universe. 
\section{dark heating}
\label{sec:darkheating}
In this Section, we will set up the basic differential equations to study the baryon and DM temperature evolution in presence of the DM viscosity and DM-gas interaction.
Considering the viscous effect of dark matter and DM-baryon interaction, the temperature of baryon and dark matter evolve as 
\begin{equation}
\frac{dT_{G}}{dz} = \frac{2T_{G}}{(1+z)} + \frac{\Gamma_{C}}{(1+z)H} (  T_{G} - T_{C}) +  \frac{2}{3(1+z)H}\frac{d{Q_{G}}}{dt}~~,
\label{eq:baryon_temp}
\end{equation}
\begin{equation}
\frac{dT_{\chi}}{dz} = \frac{2T_{\chi}}{(1+z)} +  \frac{2}{3(1+z)H}\frac{d{Q_{\chi}}}{dt} -  \frac{6}{(1+z)H}\left[ \frac{m_{\chi}\zeta_{\chi}H^{2}}{\rho_{\chi}}\right]
\label{eq:dm_temp}
\end{equation}
where $T_{G}$ is gas temperature.
Here the Compton scattering rate, $ \Gamma_{C} $ is given by
\begin{equation}
\Gamma_{C} = \frac{8\sigma_{T}a_{r} T^{4}_{C}(z)}{3m_{e}}\frac{x_{e}}{(1+x_{He} + x_{e})}~~,
\end{equation}	
where the Compton scattering cross section, $\sigma_{T} =  6.65\times 10^{-25}$cm$^2$, radiation constant $ a_{r} = 7.5657\times10^{-16} $Joule m$^{-3}$ K$^{-4}$.  Here electron fraction $ x_{e} \equiv  \frac{n_{e}}{n_{H}} $ and helium fraction,  $ x_{He} \equiv \frac{n_{He}}{n_{H}} $, where $n_{e}$, $n_{He}$ and  $n_{H}$ represents the electron, helium and hydrogen number density respectively. Also, the CMB temperature at any redshift is given by $T_{C}(z) = T_{0}(1+z)$, where $T_{0}\sim 2.72 K$.

Further, the last term $ \frac{d{Q_{G}}}{dt} $ in Eq. (\ref{eq:baryon_temp})  represents the heat transfer rate by baryon to DM due to DM-baryon interaction.
In order to have sufficient cooling of the gas at the cosmic dawn era, the interaction between the DM and gas must be sufficiently large, which can be done by considering
the DM-gas interaction of the Rutherford  type $ \sigma = \hat{\sigma}v^{-4}_{\mathrm{rel}} $, where $v_{\mathrm{rel}}$  is the relative velocity between DM and gas. For the DM-gas interaction of type $ \sigma = \hat{\sigma}v^{-4}_{\mathrm{rel}} $,
the  heat transfer rate by baryonic to DM is given by  \cite{Tashiro:2014tsa,Munoz:2015bca,Munoz:2018pzp}
\begin{equation}
\frac{d{Q_{G}}}{dt} =\sum_{I}^{} \sqrt{\frac{2}{\pi}}\frac{\mu_{I}}{(m_{I} + m_{\chi})}\left[  x_{I}\frac{e^{-r^2_{I}/2}}{u_{\mathrm{th}}^3} \right] \bigg( T_{G}(z) - T_{\chi}(z)\bigg)n_{\chi}(z)\hat{\sigma}_{I}
-\frac{\rho_{\chi}}{\rho_{M}}\mu_{I}v_{\mathrm{rel}}D(v_{\mathrm{rel}})~,
\label{eq:heat_tranfer}
\end{equation}
where  $\hat{\sigma}_{I}$ is  DM-gas scattering cross-section and $ \rho_{M} $ is total matter density. Here $ I =( H, He, e, p) $ are the species from which the DM can interacts. The different notation in Eq. (\ref{eq:heat_tranfer})  are given by
$$ \ \ \mu_{I} = \frac{m_{\chi} m_{I}}{m_{\chi} + m_{I}},  \ x_{I} = \frac{n_{I}}{n_{H}},  r_{I} = \frac{v_{\mathrm{rel}}}{u^{I}_{\mathrm{th} }} \mathrm{ \ and} \  (u^{I}_{\mathrm{th} })^2 = \frac{T_{G}}{m_{I}} + \frac{T_{\chi}}{m_{\chi}}~. $$
If DM is  interacting with the Hydrogen only then $m_{I}= m_{H}$ and  $x_{I}= 1$. The second term,  $ \frac{d{Q_{\chi}}}{dt} $ in Eq. (\ref{eq:dm_temp}) represents the  heat absorption rate by the dark matter particles, which can be obtained by $ G \leftrightarrow \chi $. 
The second term in the heat transfer rate (for both the baryon and DM) corresponds to the drag term which is small for the DM mass is less than the GeV scale. For larger DM mass ( $m_{\chi}>1$GeV), the contribution of drag term becomes large and will heat both the DM and gas \cite{Munoz:2015bca}.  
Also,
it is clear from Eq. (\ref{eq:dm_temp}) that the heating effect due to DM viscosity is prominent for large DM viscosity and large DM mass.

In the presence of DM-baryon interaction, the evolution equation of DM-baryon relative velocity is given as \cite{Munoz:2015bca,Munoz:2018pzp}
\begin{equation}
\frac{dv_{\mathrm{rel}}}{dz} =  \frac{v_{\mathrm{rel}}}{(1+z)} + \frac{D(v_{\mathrm{rel}})}{(1+z)H}
\label{eq:vel}
\end{equation}
\begin{equation}
\mathrm{where} \ \  D(v_{rel}) \equiv - \frac{dv_{\mathrm{\mathrm{rel}}}}{dt} = \sum_{I}^{}\frac{\rho_{I}}{\rho_{B}}\frac{\rho_{m}\hat{\sigma}_{I}}{m_{I} + m_{\chi}}\frac{1}{v^{2}_{\mathrm{rel}}}F(r_{I})
\end{equation}
and the function $F(r_{I}) $ is given by
\begin{equation}
F(r_{I}) \equiv \mathrm{Erf}(\frac{r_{I}}{\sqrt{2}}) - \sqrt{\frac{2}{\pi}}r_{I}e^{-r^{2}_{I}/2}
\end{equation}
The first term of  Eq. (\ref{eq:vel}) represents the dilution with the Hubble expansion whereas the second term of is a result of the interaction between DM and baryons. It is clear from the Eq. (\ref{eq:vel}) that in presence of DM-baryon interaction the relative velocity decreases fastly.

In order to solve the DM and baryon temperature evolution, we need to provide an equation for the electron fraction, $ x_e(z) $.  The electron fraction evolution can be written as \cite{AliHaimoud:2010dx}
\begin{equation}
 \frac{dx_e}{dz} = \frac{1}{H(1+z)}\frac{\frac{3}{4}R_{Ly\alpha}+\frac{1}{4}\Lambda_{2s,1s}}{\beta_{B}+\frac{3}{4}R_{Ly\alpha}+\frac{1}{4}\Lambda_{2s,1s}}\Big(n_H x_e^2\alpha_{B}(T_b)-4(1-x_e)\beta_{B}e^{-E_{21}/T_C} \Big)
 \label{electron}
\end{equation}
where $\alpha_{B}$ is the case-B recombination coefficient and given the fitting function: $\alpha_{B}=10^{-19}\frac{a t^b}{1+c t^d} m^3 sec^{-1} $, with a=4.309, b=-0.6166, c=0.6703, d=0.5300 and $t=\frac{T_B}{10^4 K}$. Photoionization correspond to this recombination rate is given by: $\beta_{B}= \frac{(2 \pi \mu_e T_C)}{4 h^3}e^{E_2/T_C}\alpha_{B}(T_B=T_C)$, $E_2=3.4 eV$. Here 
Ly$\alpha$ is the photon escape life given by: $R_{Ly\alpha}=\frac{8 \pi H}{3 n_H (1-x_e)\lambda_{Ly\alpha}^3}$, the Ly$\alpha$ rest wavelength  is $\lambda_{Ly\alpha}$ = 121.5682 nm. The H two photon decay rate is $\Lambda_{2s,1s}=8.22 sec^{-1}$ and  $E_{21}$ energy correspond to Ly$\alpha$ wavelength \cite{Seager:1999bc}\cite{Seager:1999km}. 

The evolution of baryon and dark matter temperature can be obtained by solving the coupled differential equations 
Eq.\eqref{eq:dm_temp}, Eq.\eqref{eq:baryon_temp},  Eq.\eqref{eq:vel} and Eq.\eqref{electron} with  initial conditions at the redshift $z=1010$. We assume DM temperature is zero, $T_{\chi}(1010)=0$, baryon  follow the CMB temperature  $T_G(1010)=2.35\times 10^{-10}$ GeV, electron fraction, $x_e(1010)=0.057$ \cite{Seager:1999bc} and the relative velocity $v_{\mathrm{rel}}(1010) \sim29$ Km$/sec \sim 10^{-4}$\cite{Barkana:2018qrx}. 
 
 \section{Results}
 \label{sec:results}
As we have seen in the previous Section \ref{sec:darkheating} that the presence of the dissipative effect in DM fluid cause the heat production in the dark sector and hence change the evolution history of DM temperature. Although the presence of DM-gas interaction cools the gas but the DM dissipation, if sufficiently strong,  may make the cooling of the gas inefficient.
In this Section, we will check the dependency of  DM viscosity ( $ \bar{\zeta} $), DM mass ($ m_{\chi} $) and the DM-gas interaction cross-section ($ \hat{\sigma} $) on the temperature evolution for DM and gas. Later, we see that the requirement of EDGES signal explanation put the strong constraints on the above-aforementioned quantities. 
\subsection{Evolution of DM and baryon temperature }
It is clear from the Eqs. \eqref{eq:dm_temp}, \eqref{eq:baryon_temp},  \eqref{eq:vel} and \eqref{electron}, that the gas and DM temperature depends on the DM viscosity parameters ($\gamma, \bar{\zeta} $), DM mass ($ m_{\chi} $) and the DM-gas interaction cross-section ($ \hat{\sigma} $). Below, we will study the effects of the aforementioned parameters on temperature evolution one by one. In our analysis, we will consider that the DM is interacting with the hydrogen and at last we will generalize our analysis for the other components of the gas like Helium, electron, and proton.  
\subsubsection{Temperature dependency on  DM viscosity, $\bar{\zeta}$ and DM mass, $m_{\chi}$ } 
The magnitude of the viscosity parameter after which the viscous dissipation becomes important in the dark sector can be obtained by comparing the viscous heating (third term) and baryonic heating (second term) in the Eq.(\ref{eq:dm_temp}).  To quantify the relative strength of the two competing heating mechanism of dark sector, we define a quantity, $ \Upsilon $ which is the magnitude of ratio of baryonic heating to DM viscous heating as
\begin{equation}
\Upsilon=\frac{1}{\bar{\zeta}}\frac{\hat{\sigma}_{I}m^{5}_{\chi}}{H_{0}m^{2}_{\mathrm{pl}}}\frac{\mu_{I}}{m_{I}+m_{\chi}}\left[\frac{n_{I}n_{\chi}}{ m^{5}_{\chi}H^{2}} \left( \frac{\rho_{\chi0}}{\rho_{\chi}}\right)^{-\gamma}  \left\lbrace \sqrt{\frac{2}{\pi}}  \frac{x_{I}e^{-r^{2}_{I}/2}}{u_{\mathrm{th}}^3} \big|T_{\chi}- T_{G} \big|
+\frac{m_{I}F(r_{I})}{v_{\mathrm{rel}}}\right\rbrace \right] 
\end{equation}
where $m_\mathrm{pl}=\frac{1}{\sqrt{G}}$ is the reduced planck mass. The term inside the square bracket  decreases with the cosmic evolution and the term before the square bracket is independent on the redshift and its value  will decide the strength of $\Upsilon$. 
The condition that the viscous heating of DM will prominent over the baryonic heating at redshift $z$ is $ \Upsilon\ll1 $ and the DM viscous heating will be unimportant over the baryonic heating when $ \Upsilon\gg1 $.
\begin{figure}%
	\subfigure[]{%
		\label{fig:tem-a}%
\includegraphics[height=2in,width=3.6in]{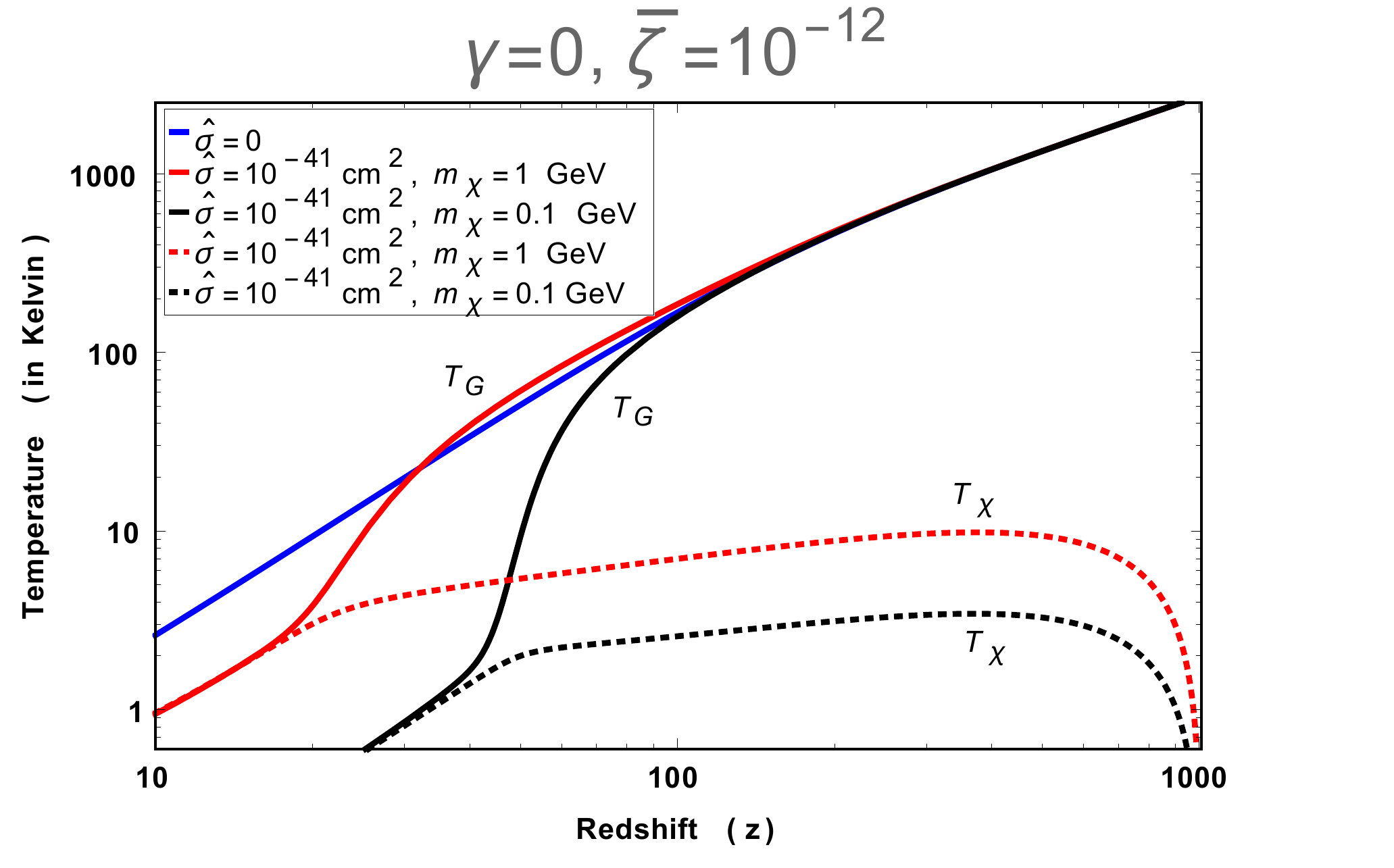}}%
	\subfigure[]{%
		\vspace{-0.3cm}%
		\label{fig:tem-b}%
\includegraphics[height=2in,width=3.6in]{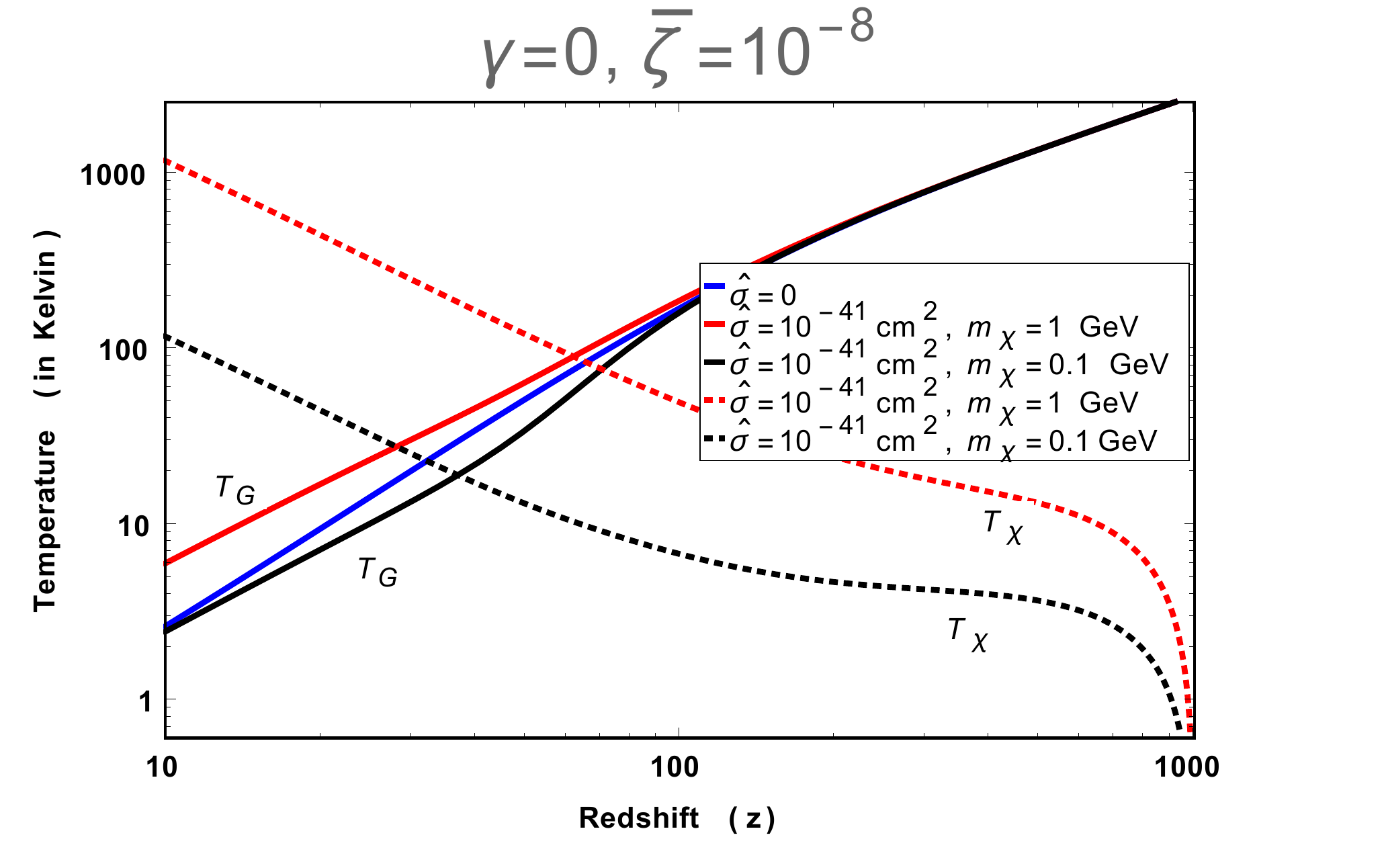}} \\
	\vspace{-0.3cm}%
	\subfigure[]{%
		\label{fig:tem-c}%
\includegraphics[height=2in,width=3.6in]{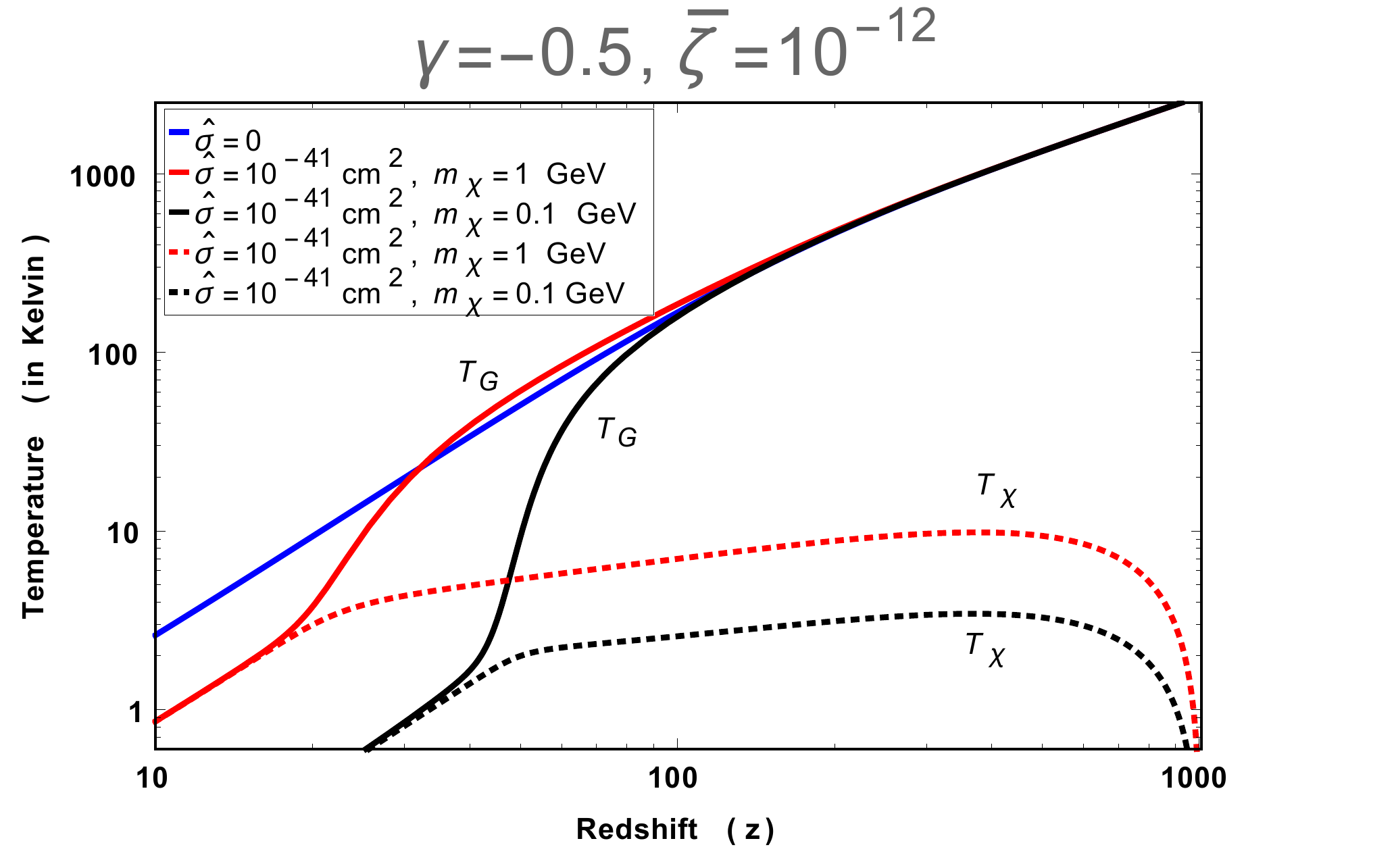}}%
	\subfigure[]{	\vspace{-0.3cm}%
		\label{fig:tem-d}%
\includegraphics[height=2in,width=3.6in]{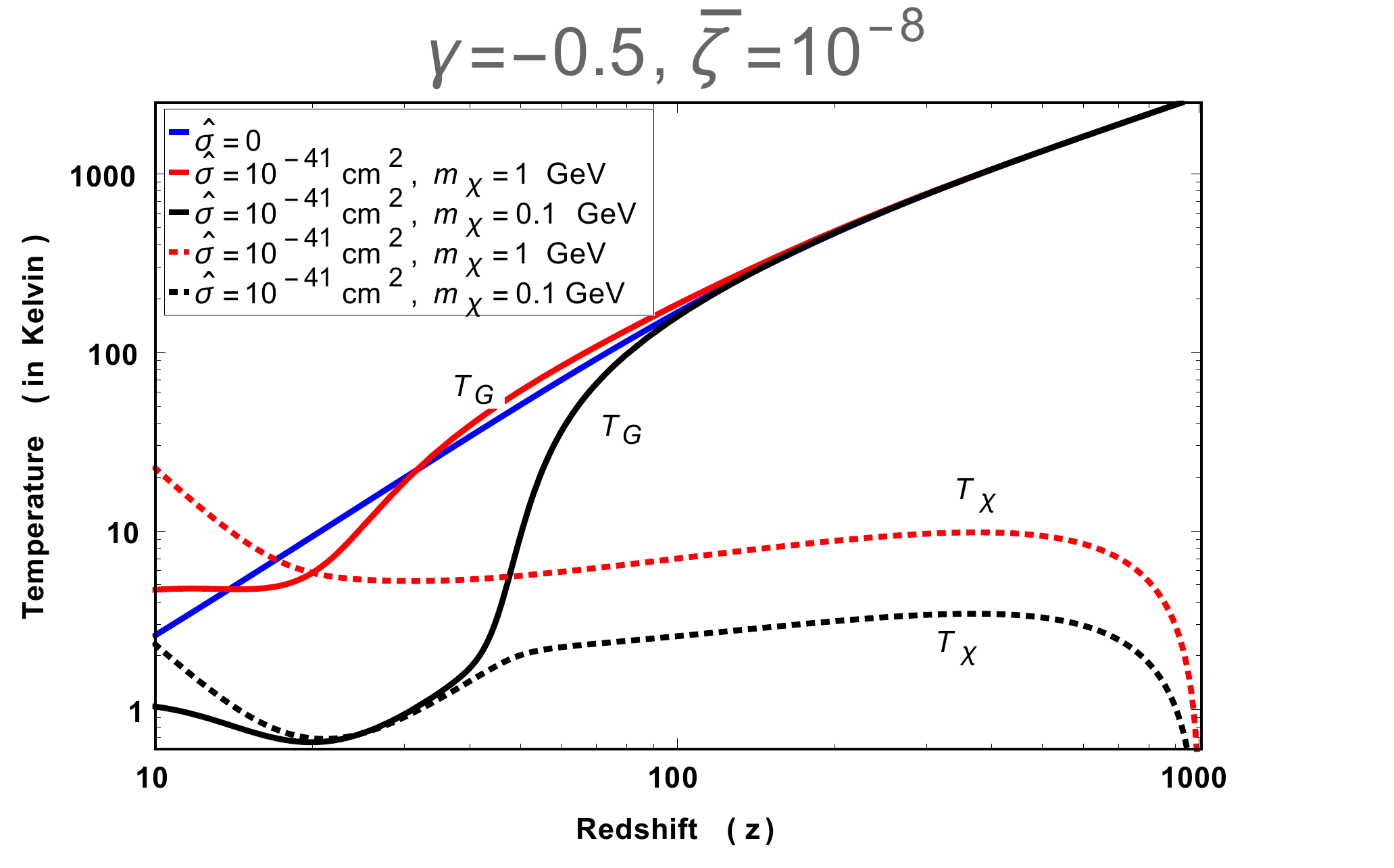}}%
\caption{Temperature evolution of baryon and dark matter vs redshift ($z$) for different viscosity parameter, DM mass, and DM baryon cross-section. The solid blue line in all figures corresponds to $\hat{\sigma}=0$,  i.e. no interaction between dark matter and baryon also no viscosity in DM. The red and black solid curves correspond to gas temperature whereas the red-dashed and black-dashed curves corresponds for dark matter temperature. Fig. \ref{fig:tem-a} and \ref{fig:tem-b} corresponds to constant viscosity parameter: $ \bar{\zeta}=10^{-12}$ and $ \bar{\zeta}=10^{-8}$  which remain constant with the redshift, i.e. $\gamma=0$.  Also, Fig. \ref{fig:tem-c} and \ref{fig:tem-d} correspond to viscosity parameter: $ \bar{\zeta}=10^{-12}$ and $ \bar{\zeta}=10^{-8}$  which varries with the redshift, i.e.  $\gamma=-1/2$.}
	\label{temp_evol}%
\end{figure} 

Fig.\eqref{temp_evol} shows the temperature evolution of dark matter and baryons. The upper  Figures \ref{fig:tem-a} and \ref{fig:tem-b} are temperature plot for which the viscosity parameters are constant with redshift, $\gamma=0$  and lower Figures \ref{fig:tem-c} and \ref{fig:tem-d} are  temperature plot for which the viscosity parameters varies with the redshift, i.e. $\gamma=-0.5$.  In all the figures, the solid blue line corresponds to the evolution of baryon temperature when the interaction between the dark matter and baryon is zero and also DM viscosity is zero. 
In Fig. \ref{fig:tem-a}, we see that for viscosity parameter,   $\bar{\zeta}=10^{-12}$, both the dark matter and baryon temperature evolution follow the same pattern as there is no viscosity in dark sector (see Fig.1 of Ref.\cite{Munoz:2015bca}). This implies that such a small DM viscosity does not produce the sufficient dissipation that changes the DM and baryon temperature. In this case, throughout the cosmic evolution $ \Upsilon\gg1 $, hence the heating corresponding to DM-baryon interaction (second term in right-hand side) dominant over the viscous dissipation (third term in right-hand side) in the Eq.\eqref{eq:dm_temp}.
But for higher viscosity parameter, $  \bar{\zeta}=10^{-8} $, the DM dissipation becomes prominent and hence increase both the DM and gas temperature, see Fig. \ref{fig:tem-b}.  At this time $ \Upsilon\ll1 $ hence the heating due to the large viscosity of dark matter start dominating over the heating due to the DM-baryon interaction and thus the DM temperature increases sharply and becomes greater than the baryon at low redshift. 

A similar argument can be given for the varying viscosity case, which is plotted in the lower  Figures \ref{fig:tem-c} and Fig. \ref{fig:tem-d}. But in this case, the value of viscosity parameter up to which the DM and gas follow same temperature evolution and behaves like as there is no viscosity in DM is slightly larger, $  \bar{\zeta}=10^{-11} $.
Meanwhile, we also point out that the DM temperature corresponds to constant viscosity parameter,  $\gamma=0,\bar{\zeta}=10^{-8} $ (i.e. Figure  \ref{fig:tem-c}) is larger than the varrying viscosity parameter, $\gamma=-0.5,\bar{\zeta}=10^{-8} $ (i.e. Figure  \ref{fig:tem-d}) at small redshift. This happens because throughout the cosmic eolution history the dissipation correspond to the constant viscosity parameter is larger in comparison with the varrying viscosity.

Here we also see the effect of mass variation onto the temperature evolution of DM and gas. In the upper Figures  \ref{fig:tem-a} and  \ref{fig:tem-d}, we find that while increasing the DM mass, from $0.1$ to $1$ GeV, the contribution from the viscous dissipation becomes large and hence increase the DM temperature and also the gas temperature. 
Thus we conclude that while increasing the DM viscosity and DM mass (large DM mass range, i.e. $0.1$ to $1$ GeV) cause more heating of DM and baryon. Hence in order to cool the gas efficiently, the DM mass and DM viscosity should not be very large.  

Until now in this subsection, we have only seen the effect of viscosity on the temperature evolution of the gas and DM. We have not given the interpretation of the magnitude of the viscosity parameters, which are severely constrained from the observation.  In Ref. \cite{Velten:2013pra}, authors have put a strong constraint on the viscosity parameters from the structure formation. There they argued that in order to form non-linear structures, like dwarf galaxies, the viscosity $\bar{\zeta}\leq 5\times 10^{-11}$, otherwise, the larger viscosity will decrease the growth of inhomogeneities and wash out the structures. 
The contribution from this viscosity, $\bar{\zeta}\leq 5\times 10^{-11}$ is not sufficient to increase the gas temperature efficiently and hence does not affect the 21 cm absorption signal.
Here in our analysis, we consider the larger values of viscosity to show how the higher dissipation can affect the temperature evolution.

\subsubsection{Temperature dependency on  DM-baryon interaction cross-section $ \hat{\sigma}$} 
\begin{figure}[]
\begin{minipage}[]{0.45\linewidth}
		\centering
\includegraphics[width=\linewidth]{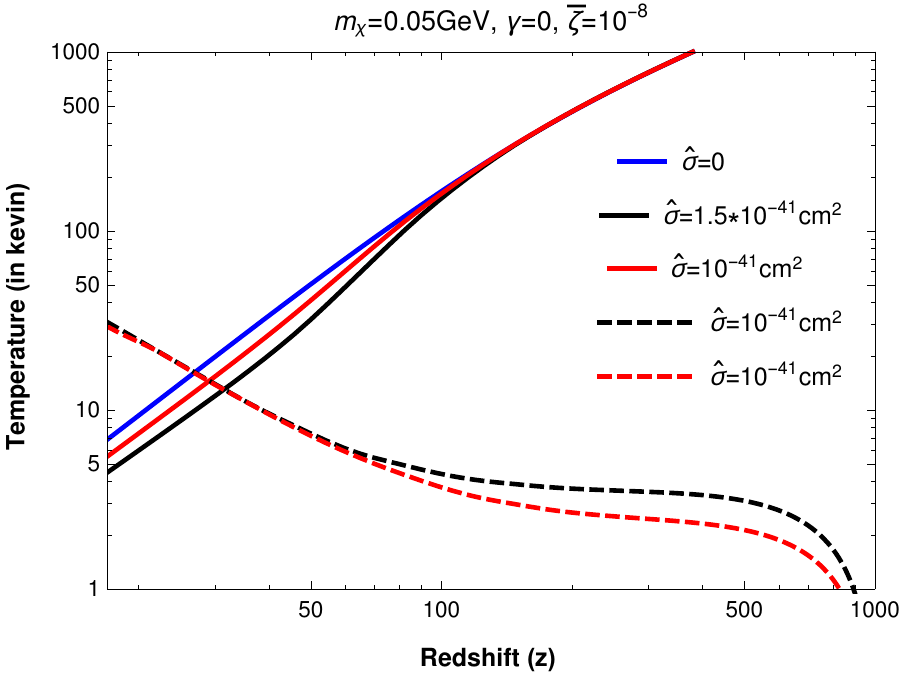}
\caption{The temperature of DM and gas at constant viscosity and DM mass as a function of redshift for different values of  DM-gas scattering cross-section, $\hat{\sigma}$. As the  $\hat{\sigma}$ increase,  the gas temperature $T_{G}$ decreases but the DM temperature, $T_{\chi}$ increases, throughout the cosmic evolution.} 
		\label{fig:sigma}
	\end{minipage}
	\hspace{0.2cm}
\begin{minipage}[]{0.48\linewidth}
		\centering
\includegraphics[width=\linewidth]{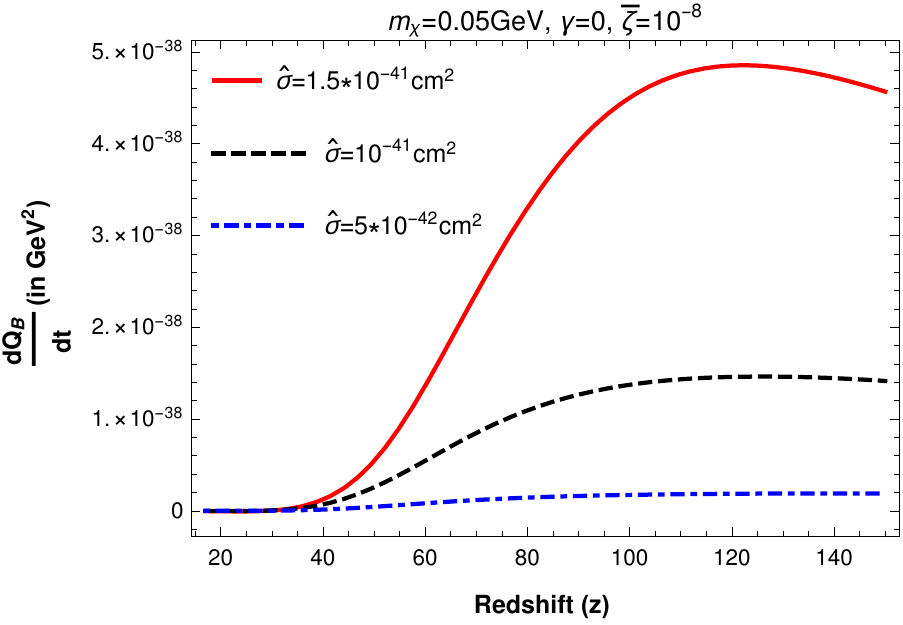}
\caption{The heat transfer rate by the gas to DM has been plotted as a function of the redshift for different values of DM-gas scattering cross-section, $\hat{\sigma}$. As the $\hat{\sigma}$ increases the heat transfer rate increases at the earlier time and decreases at the late time (low redshift).  } 
		\label{fig:QdotG}
	\end{minipage}
\end{figure}
In Fig. \ref{fig:sigma}, we plot the temperature of gas as the function of redshift for different values of cross-section, $\hat{\sigma} $.  For this purpose, we fix the  viscosity parameter, $\gamma=0$, $\bar{\zeta}= 10^{-8}$ and DM mass, $m_{\chi}=0.05$ GeV. Here the blue line represents the gas temperature when there is no interaction between the DM-gas and also no viscosity in the dark sector. 
The interaction between the DM and gas cool the gas by transferring the energy from the gas to DM. We see that for $\hat{\sigma}= 10^{-41} $ cm$^{2}$ (red thick line), the baryon temperature at $z=17$ is $\sim 5.4$K, which is small in comparison with the standard cosmological prediction. But as the $\hat{\sigma}$ increase from the $\hat{\sigma}= 10^{-41} $ cm$^{2}$ to $\hat{\sigma}= 1.5\times10^{-41} $ cm$^{2}$, due to larger interaction between the DM and gas, the effective cooling of the gas and also the heating of the DM starts comparatively at earlier times, which in turn reduces the gas temperature  $\sim 4.4$K at redshift  $z=17$. 

We also find at some point of cosmic evolution, the DM temperature becomes larger than the gas temperature and can start heating to the gas and hence increase the gas temperature at low redshift. But we find that in this mass range the heating to gas is not very effective. We checked that for the larger DM mass, increasing the $\hat{\sigma}$ causes more energy transfer from the gas to DM at the earlier time and when $T_{\chi}>T_{G}$, the heating of gas becomes prominent. Thus at low redshift, the gas temperature increases fastly for large DM mass in comparison with the small DM mass.

In order to understand the cooling of gas due to the large $\hat{\sigma}$ values, we also plot the heat tranfer rate by the gas, $ \frac{d{Q_{G}}}{dt} $ given in Eq. (\ref{eq:heat_tranfer}),  as a function of redshift for the different values of the DM-gas interaction cross-section $\hat{\sigma}$ in Fig. \ref{fig:QdotG}. Here we take the same values of parameters, which we have taken Fig. \ref{fig:sigma}.  We find that  for small $\hat{\sigma} $ value, i.e. $\hat{\sigma}= 5\times10^{-42} $ cm$^{2}$ (blue dotted line), the heat transfer rate is small but as the $\hat{\sigma}$ increases from  $\hat{\sigma}=5\times10^{-42}$cm$^{2}$ to  $\hat{\sigma}=1.5\times10^{-41}$cm$^{2}$, the $\frac{d{Q_{G}}}{dt} $ becomes large at earlier time. Consequently, it is clear from the Eq. (\ref{eq:baryon_temp}) that the cooling of the gas for large $\hat{\sigma}$ is comparatively larger and $T_{G}$ becomes small at the $z=17$. Although at small redshifts, when $T_{\chi}>T_{G}$,  $\frac{d{Q_{G}}}{dt} <0$ but at that redshift the heating effect is not sufficient enough to increase the gas temperature.
%
\begin{figure}
	\centering
\includegraphics[width=0.65\linewidth,height=0.45\linewidth]{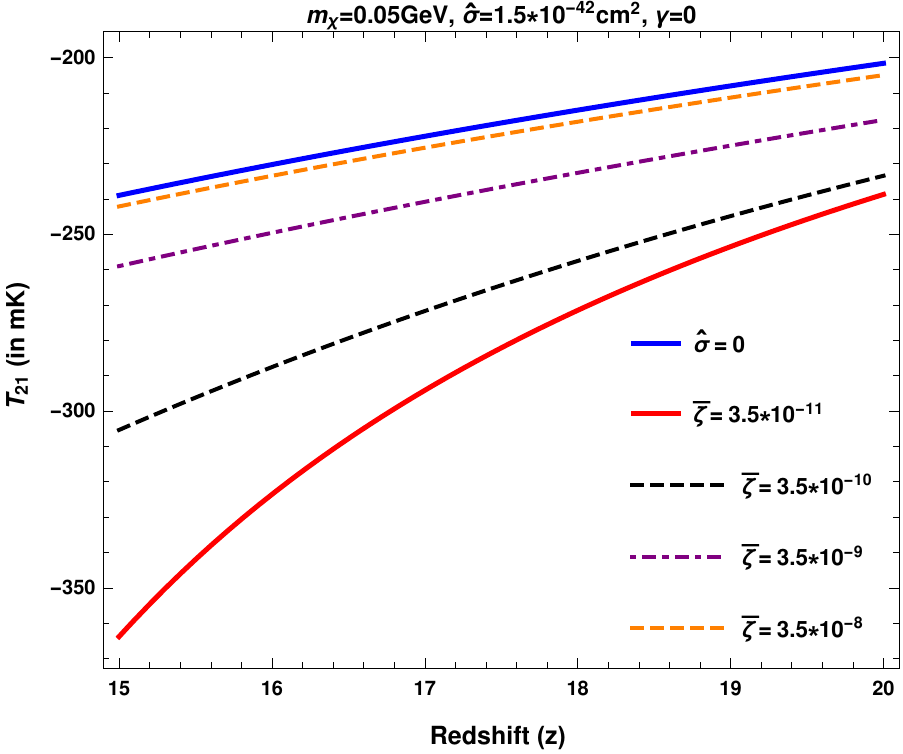}
\caption{The plot of brightness temperature, $T_{21}$ as a function of redshift for the different values of viscosity. The blue line represents the standard scenario with no DM-Hydrogen interaction ($\hat{\sigma}=0 $) and also no viscosity. For the other curves $ \hat{\sigma} =5\times 10^{-42}$ cm$^{2}$ and viscosity is constant, $\gamma=0$. As the DM viscosity increases, the strength of absorption signal decreases, i.e brightness temperature, $T_{21}$ increases. }
	\label{fig:bright}
\end{figure}
\subsection{Brightness Temperature}
In order to see the effect of the viscosity on the brightness temperature, as given in Eq.(\ref{eq:heat_tranfer}), we plot brightness temperature, $T_{21}$ as a function of redshift for the different values of viscosity in Fig. \ref{fig:bright}. Here we plot the $T_{21}$ at small redshift $ 15\leq z \leq 20 $, because at these redshifts the spin temperature couples with the gas temperature, i.e. $T_{S}=T_{G}$ in Eq. (\ref{eq:brightness}). For this purpose, we fix the $\hat{\sigma}=1.5\times 10^{-41}$cm$^{2}$ and DM mass $m_{\chi}=0.05$GeV.  Here the blue thick line represents the brightness temperature when there is no interaction between the DM and gas and also no viscosity in DM. The other lines show the behaviour of brightness temperature in presence of DM-gas interaction and DM viscosity.

It is evident from Fig. \ref{fig:bright} that when the viscosity is small, $\bar{\zeta}=3.5\times 10^{-11}$ (red line), the gas can cool efficiently via DM interaction and make the brightness temperature $T_{21}$ more negative and explain the EDGES observation.  Furthermore, as the viscosity increases, from the $\bar{\zeta}=3.5\times 10^{-10}$ to $\bar{\zeta}=3.5\times 10^{-9}$,(black dashed and purple dot dashed lines) the gas temperature start increasing and consequently diminishes the absorption signal. For sufficient large viscosity parameter, $\bar{\zeta}=3.5\times 10^{-8}$, (orange dashed line) the signal strength becomes very small. This implies that in order to explain the EDGES signal, the DM viscosity should not very large.
\begin{figure}
	\centering
\includegraphics[width=0.7\linewidth,height=0.45\linewidth]
{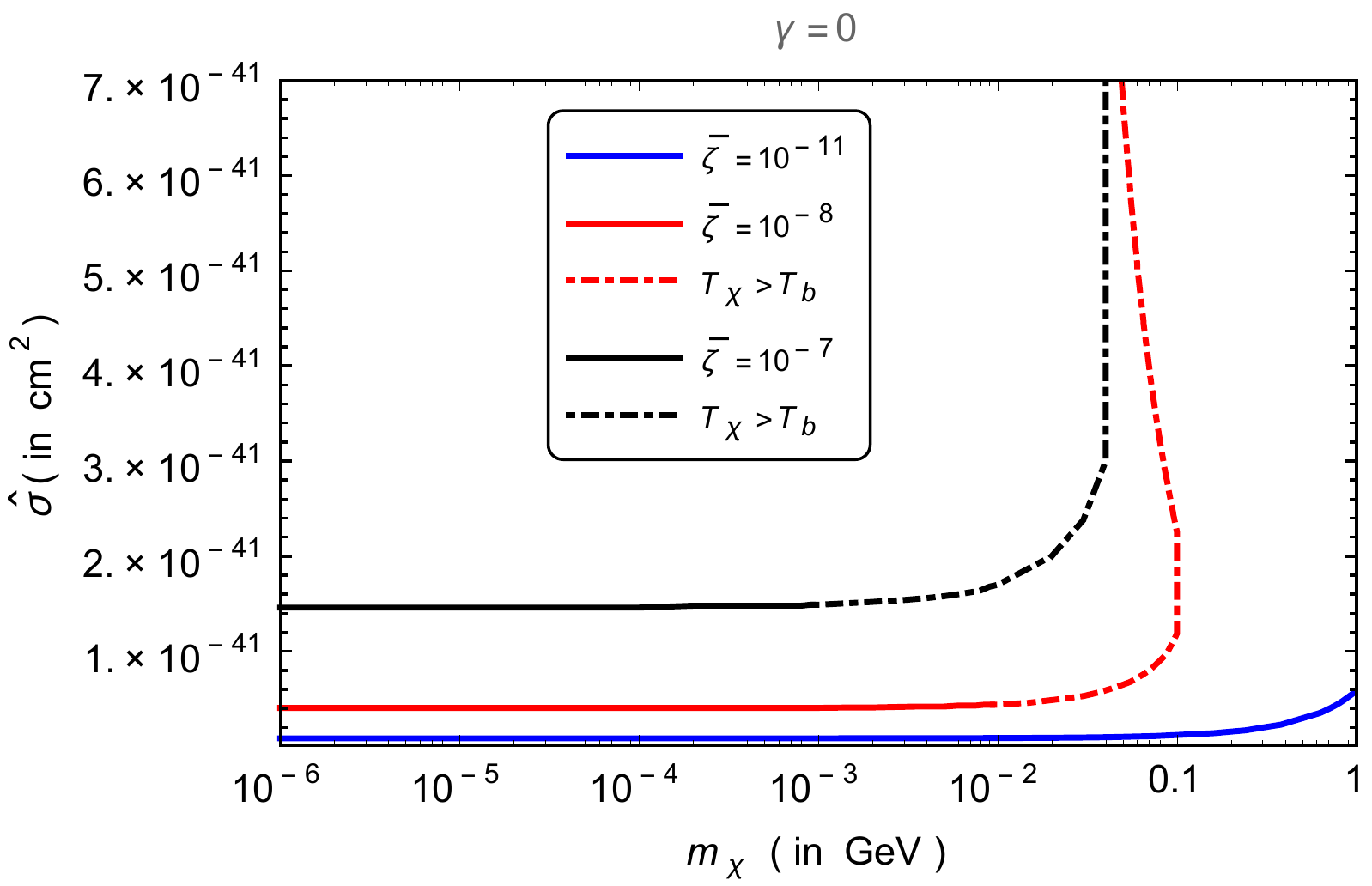}
\caption{The DM-Hydrogen scattering cross section vs DM mass needed to fit the EDGES absorption signal are plotted for three constant viscosity parameters $\bar{\zeta}$=$10^{-11}$ (blue line), $10^{-8}$ (red line) and $10^{-7}$(black line). The solid lines corresponds to $T_{\chi}<T_{G}$ and dot dashed lines corresponds to $T_{\chi}> T_{G}$. The cross-section corresponds to the brightness temperature $T_{21}=-300$mK, which is the upper limit of the EDGES observation.} 
	\label{fig:constant_vis}
\end{figure} 
\subsection{Constraints on DM-baryon scattering cross-section and DM mass }
\label{sub:csmass}
In the previous subsections, we have analyzed the effect of different parameters on DM viscosity and below we will use this knowledge to constrain those parameters.
In order to get the EDGES absorption signal at $z=17$, i.e. that satisfy $T_{G}=5.20 K$,
we get a range of parameter for DM-baryon scattering cross section as a function of dark matter mass. 
 
In Fig.\eqref{fig:constant_vis}, we plot DM-baryon scattering cross section vs DM mass for three viscosity parameters $\bar{\zeta}=10^{-11}$, $\bar{\zeta}=10^{-8}$ and $\bar{\zeta}=10^{-7}$ which are constant ($\gamma=0$) with the redshift.
Here all lines correspond to the gas temperature $T_{G}=5.20 K$ at $z=17$, which explain the upper limit of the EDGES absorption signal. Further the solid lines corresponds to $T_{\chi}<T_{G}$ and dot dashed lines corresponds to $T_{\chi}>T_{G}$. We find that to explain the EDGES observation, the maximum limit for the DM mass corresponds to  DM viscosity parameter, $\bar{\zeta}=10^{-11}$  is same as ideal DM case but this time the cross section has bee increased comparatively.
Furthermore, as we increase the DM viscosity from $\bar{\zeta}=10^{-11}$ to $\bar{\zeta}=10^{-7}$, the DM temperature increases which oppose the heat transfer from the baryonic matter.
So the required baryon temperature at $z=17$ necessary for EDGES observation, cannot be achieved with the above given mass and cross section.
To obtain  $T_{G}=5.20\, $K (upper limit of EDGES signal) for the same viscosity parameter, $\gamma=0,\, \bar{\zeta}=10^{-8}$, either we need to decrease the DM mass or increase the DM-baryon interaction or the combination of both.
It is clear that while increasing the viscosity parameter, $ \bar{\zeta} $ from $10^{-11}$ to $10^{-7}$ the  DM-baryon scattering cross section, $\hat{\sigma}$ has been increased from $8.5\times10^{-43}$cm$^{2}$  to $1.5\times10^{-41}$cm$^{2}$ and also DM mass range has been decreased.

In Fig.\eqref{fig:constant_vis}, we also find that for higher viscosity of dark matter ($\bar{\zeta}>10^{-11}$), the DM temperature also increase above the baryon temperature (here $T_{G}=5.20$K) after a certain DM mass range. The red dot dashed and blue dot dashed line in Fig.\eqref{fig:constant_vis} corresponds to  $T_{\chi}> T_{G}$ for $\bar{\zeta}=10^{-8}$ and $10^{-7}$ respectively.
As the DM temperature increases above the baryon temperature, it may heat the baryon due to DM-baryon coupling and
hence can erase the excess dip in 21 cm signal. Hence in order to cool the gas further the DM-gas interaction cross-section must be increased. The DM mass scale after which the gas heating and hence $\hat{\sigma}$ increases sharply puts
an upper bound on dark matter mass for different viscosity parameters. The upper limit on DM mass is $0.1$ GeV and $0.02$ GeV for $\tilde{\zeta}=10^{-8}$ and $10^{-7}$ respectively.
\begin{figure}
	\centering
	\includegraphics[width=0.7\linewidth,height=0.45\linewidth]
{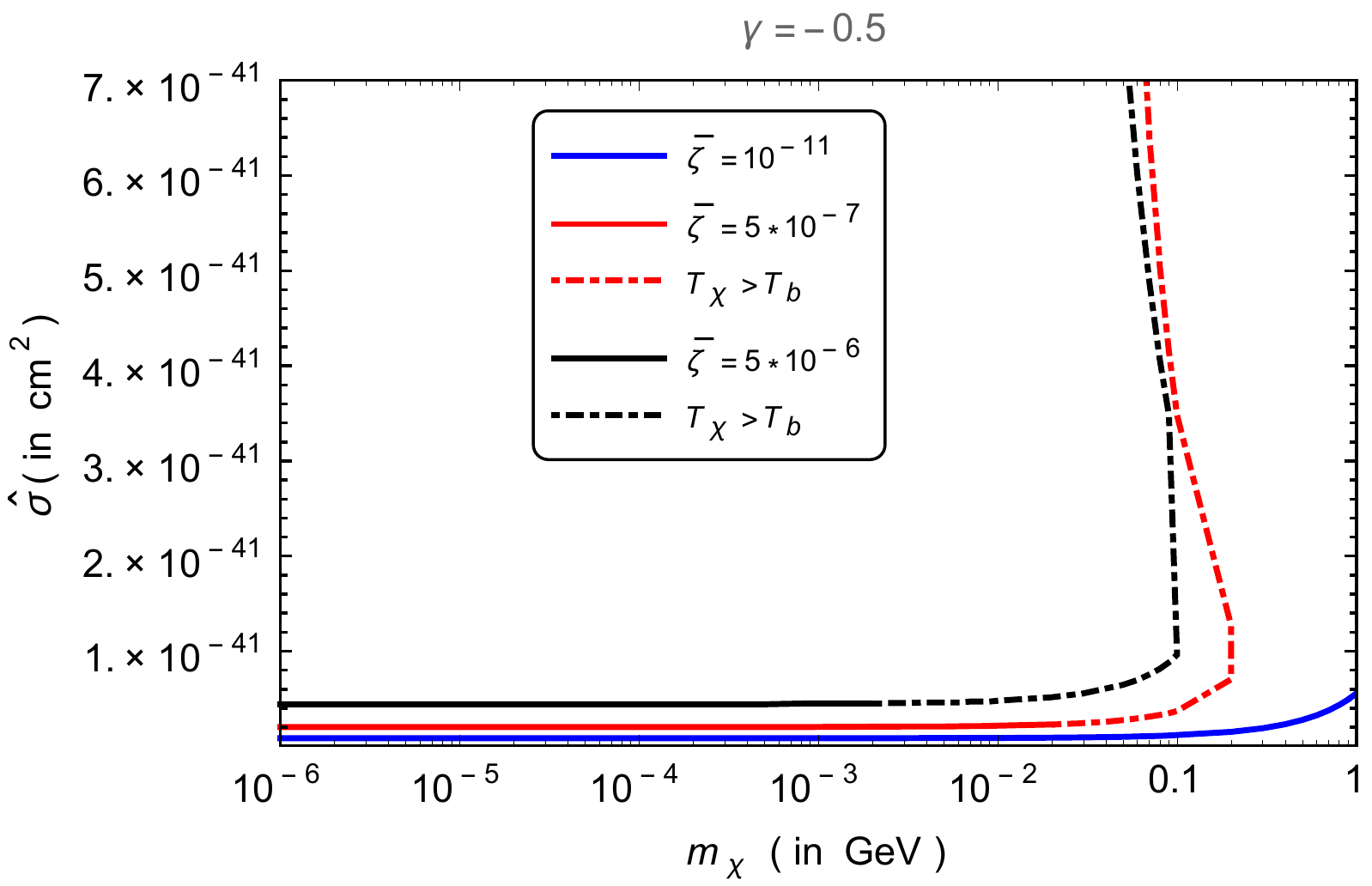}
\caption{The DM-Hydrogen scattering cross section vs DM mass needed to fit the EDGES absorption signal are plotted for three varying viscosity parameters $\bar{\zeta}$=$10^{-11}$ (blue line), $5\times10^{-7}$ (red line) and $5\times10^{-6}$ (black line). The solid lines corresponds to $T_{\chi}<T_{G}$ and dot dashed lines corresponds to $T_{\chi}> T_{G}$. The cross-section corresponds to the brightness temperature $T_{21}=-300$mK, which is the upper limit of the EDGES observation.} 
\label{fig:var_vis}
\end{figure}

Similarly, in Fig.\eqref{fig:var_vis} we plot DM-baryon scattering cross section vs DM mass for three different viscosity parameters,  $\bar{\zeta}$=$10^{-11}$ , $5\times10^{-7}$  and $5\times10^{-6}$, which varries with the redshift ($\gamma=-1/2$). Here also all lines represents the gas temperature $T_{G}=5.20$K at $z=17$, 
in which the solid line corresponds to $ T_{\chi}<T_{G} $ and dot dashed line corresponds to $ T_{\chi}>T_{G} $.  
Here we also find that to explain the EDGES observation, the maximum limit for the DM mass corresponds to DM viscosity parameter, $\bar{\zeta}=10^{-11}$  is same as ideal DM case but the cross section has been increased.
Also while increasing the viscosity parameter, $ \bar{\zeta} $ from $5\times 10^{-11}$ to $5\times10^{-6}$, in order to keep the baryon temperature at $5.20 K$, the DM-baryon scattering cross section, $\hat{\sigma}$ increases from $8.5\times10^{-43}$cm$^{2}$ to $ 4.42\times 10^{-42}$ cm$^{2}$. 
In this case the upper limits on DM masses are $0.2$ GeV and $0.1$ GeV corresponds to viscosity parameters $5\times 10^{-11}$ to $5\times10^{-6}$ respectively. 	

In above analysis, we point out that for a constant viscosity parameter, the value of DM-baryon interaction cross-section, $\hat{\sigma}$ obtained for DM masses less than $0.1$ GeV and $0.01$ GeV (i.e. $4\times10^{-42}$cm$^{2}$ for $m_{\chi} < 0.1$ GeV and $1.5\times10^{-41}$cm$^{2}$ for $m_{\chi}<0.01$ GeV) 
are the within the $95$\% confidence limit of $\hat{\sigma}$ obtain from the Planck temeprature, polarization and lensing measurements \cite{Boddy:2018wzy}.
If we further increase the viscosity parameter then it reduces the DM mass range and increases $\hat{\sigma}$ and hence the $\hat{\sigma}$ might be greater than the maximum limit as discussed in Ref. \cite{Boddy:2018wzy} and hence put an upper limit on the viscosity parameter.
Here we find that the for constant viscosity, $\gamma=0$ the maximum limit on the viscosity is $\bar{\zeta}\sim 10^{-7}$ which is larger by $\sim 4$ orders of magnitude in comparison with the limit obtained from the structure formation. For variable viscosity case ($ \gamma=-1/2 $), the maximum limit on viscosity is approximately two times larger in comparison with the constant viscosity case. This limit will be further improved if one does a more precise calculation of the temperature evolution of gas.

In our analysis, we have taken the upper bound of EDGES absorption signal, i.e. $T_G=5.20$K to get the parameter space for $\hat{\sigma}$ and DM mass. If we consider the lower bound of the reported EDGES signal, i.e. $T_G=1.68$K, we need to increase the baryon-DM interaction for a fixed viscosity parameter. Thus in Fig. \eqref{fig:constant_vis} and \eqref{fig:var_vis}, all the curvey will shift upward in the $\hat{\sigma}$-axis.

Also throughout our analysis, we consider the interaction of the DM with the hygrogen in order to derive the constraints on the DM viscosity, ($ \bar{\zeta} $) DM mass ($m_{\chi}$)  and DM-gas interaction cross-section, $ \hat{\sigma} $. But instead of hydrogen if DM interacts with the other species in the gas like helium, electron or proton, then the constraints on above parameters will change. In the approximation $ m_{\chi}\ll 1 $ GeV and cooling energy $\dot{Q}_{I}$ to be fixed, the DM-gas interaction cross-section $ \hat{\sigma}^{I} $ of these quantities will increase in comparision with the hydrogen $ \hat{\sigma}^{H} $ by \cite{Barkana:2018qrx}
\begin{equation}
\frac{\hat{\sigma}^{I}}{\hat{\sigma}^{H}}=\frac{1}{x^{I}}\left(\frac{m_{I}+m_{\chi}}{m_{p}+m_{\chi}} \right)^{2} \left( \frac{m_{p}}{m_{I}}\right)^{\frac{5}{2}}
\end{equation}
where $I=e, He, p$. Within the time of our interest $ 15\leq z\leq200 $, we can consider $ x_{\mathrm{He}}\sim\frac{1}{13} $ and $ x_{\mathrm{e}}\sim 10^{-4} $.
Considering the DM mass as an order of sub GeV, for example $m_{\chi}=0.05$ GeV, we get $ \hat{\sigma}^{He} \sim 6 \hat{\sigma}^{H}$. For the other species of gas like electron and proton, the scattering cross-section with the DM is sufficiently large in comparison with the $\hat{\sigma}^{H}$. The increment of the cross-section suggests that in order to explain the EDGES absorption signal, the maximum limit on the DM viscosity will decrease in comparison with the hydrogen. 
\section{Conclusion}
\label{sec4}
The EDGES detection of 21 cm absorption signal in the cosmic dawn era opens a new and unique window to study and test the DM properties. In standard cosmology, the DM is assumed to be ideal but if DM is viscous fluid then it may affect the EDGES observation. In this work, we studied the dissipative properties of the dark sector by assuming that dark heating has not erased the EDGES global signal. 

Considering dark matter as a viscous fluid and using the FLRW metric, we estimate the entropy production by DM fluid in the expanding Universe. The entropy production leads to heat generation in the dark sector and changes the temperature history of DM.
Later, assuming the bulk  viscosity of the form, $ \zeta_{\chi} = \zeta_{0}\left( \frac{\rho_{\chi}(z)}{\rho_{\chi_{0}}}\right)^{\gamma} $, we calculate the temperature evolution of viscous dark matter in presence of the DM-gas interaction. We then check the dependency of the gas temperature on the viscosity parameter, DM mass and  DM-gas interaction.
 
The viscosity of DM causes the heating but we show that in case of small bulk viscosity ($\bar{\zeta} \sim 10^{-12}$ for constant and $\bar{\zeta} \sim 10^{-11 }$ for varying with the redshift), the DM dissipation does not play the significant role in temperature evolution and DM behaves like as an ideal fluid. For large DM viscosity, the DM dissipation becomes prominent and increase the temperature of the DM as well as baryonic fluid. In case of sufficiently large viscosity, $\bar{\zeta} \sim 10^{-8}$ the DM temperature rise sharply and becomes larger than the baryon temperature on small redshift. The heating also depends on the DM mass and increases at higher $m_{\chi}$ values (i.e. $0.1$ to $1$ GeV). The cooling of the gas can be done by increasing the DM-gas interaction cross-section, $\hat{\sigma}$.
We thus find that for a fixed viscosity parameter, the gas temperature can be reduced via either the decreasing the DM mass ($m_{\chi}$) or increasing the DM-gas interaction cross-section ($\hat{\sigma}$).

In order to explain the EDGES absorption signal, means the temperature of baryons should not increase above to $ 5.20 $K, puts the stringent constraints on the DM viscosity ($ \bar{\zeta}$), DM mass ($ m_{\chi}$) and DM-baryon scattering cross section ($ \hat{\sigma} $). We find that the low viscosity, $\bar{\zeta}=10^{-12}$ is not enough to heat the DM and baryons hence constraints on DM mass and DM-baryon scattering cross section was same as the DM had no viscosity.  Further, on increasing the DM viscosity causes more heating in the dark sector and also in the baryonic sector through DM-baryon scattering. Consequently, to maintain the baryon temperature of $ 5.20 $K, the DM-baryon scattering cross section will increase and the mass of the DM particle will decrease.
In the case of constant DM viscosities,  $\bar{\zeta}=10^{-8}$ and  $\bar{\zeta}=10^{-7}$, the upper limit on DM masses are $0.1$ GeV and $0.02$ GeV respectively.  For varrying DM viscosity with the cosmic evolution, for viscosity parameters $\bar{\zeta}=10^{-8}$ and $\bar{\zeta}=10^{-7}$, the upper limits on DM masses are increased and given by $0.2$ GeV and $0.1$ GeV respectively.

The important result of our work is that the EDGES observation can allow large viscosity in the dark sector which is many orders of magnitude larger than the maximum limit reported from the structure formation. For a constant viscosity case, the maximum viscosity $\bar{\zeta}=10^{-7}$ and for varying viscosity case. These limits will be further improved if one does the more precise calculation of the temperature evolution of gas.
\section{Acknowledgements}
We would like to thanks Namit Mahajan for providing useful discussions and comments. AKM would also like to thank Richa Arya for the fruitful discussions and suggestions. 
ACN thanks Pratim Roy and Tripurari Srivastava for the suggestion in Mathematica program.   


\end{document}